\documentclass[letterpaper,twoside,twocolumn,english,twoside,twocomlumn,prl,aps,amssymb,showpacs,superscriptaddress]{revtex4}
\usepackage[T1]{fontenc}
\usepackage[latin1]{inputenc}
\usepackage{verbatim}
\usepackage{amsmath}
\usepackage{graphicx}
\usepackage{amssymb}

\makeatletter
\usepackage[T1]{fontenc}
\usepackage{lmodern}

\usepackage{babel}
\makeatother
\begin{document}

\title{From density to interface fluctuations: \\
the origin of wavelength dependence in surface tensions }

\author{Thorsten Hiester}

\address{Institut f\"ur Theoretische Physik, Universit\"at Erlangen-N\"urnberg, Staudtstrasse 7, D-91058 Erlangen, Germany}

\begin{abstract}
\newcommand{\gmd}{\gamma_{\mathsf{MD}}(q)}
The height-height correlation function for a fluctuating
interface between two coexisting bulk phases is derived
by means of general equilibrium properties of the corresponding
density-density correlation function. A wavelength-dependent
surface tension $\gamma(\mathbf{q})$ can be defined
and expressed in terms of the direct correlation function
$c(\mathbf{r},\mathbf{r}')$, the equilibrium density
profile $\rho_{\circ}(\mathbf{r})$ and an operator
which relates density to surface configurations. Neither
the concept of an effective interface Hamiltonian nor
the difference in pressure is needed to determine the
general structure of the height-height correlations
or $\gamma(\mathbf{q})$, respectively. This result
generalizes the Mecke/Dietrich surface tension $\gmd$
(Phys.\,Rev.\,E {\bf 59}, p.\,6766 (1999)) and modifies
recently published criticism concerning $\gmd$ (P.
Tarazona, R. Checa and, E.Chac\'{o}n: Phys.\,Rev.\,Lett. {\bf 99}, p.\,196101 (2007)).
 
\end{abstract}



\maketitle
\newcommand{\gmd}{\gamma_{\mathsf{MD}}(q)}
\newcommand{\rd}{X}
\newcommand{\rs}{\mathbf{s}}
\newcommand{\lop}{\mathsf{L}}
\newcommand{\lopd}{\mathbf{L}}
\newcommand{\lopds}{L}
Thermally excited capillary waves on the fluid interface
between two phases have drawn the attention of many
theoretical \cite{vanderWaals:657(1894),Rowlinson:1982,Buff:621(1965), Napiorkowski:1836(1993), Blokhuis:397(1999),Mecke:6766(1999)}
and experimental \cite{Benjamin:407(1997), Fradin:871(2000), Mora:216101(2003), Li:136102(2004), Luo:4527(2006), Li:235426(2006), Lin:106(2005),Royall:679(2007),Shpyrko:245423(2004)}
physicists for more than hundred years. Two approaches
have been developed in the last decades in order to
resolve the structural properties of the transition
region and its thermally driven fluctuations: Density
functional theory (DFT) is applied successfully in
order to describe the dependence of interfacial properties
on, e.g., temperature or intermolecular forces, respectively
(see, e.g.,~\cite{Telo da Gama:1091(1980)}). In particular,
the spatial dependence of the one-particle equilibrium
density $\rho_{\circ}(\mathbf{r})$, which exhibits
a smooth transition between two coexisting phases,
can be calculated within a DFT framework.

On the other hand, capillary-wave theory (CWT), see
Ref.\,\cite{Buff:621(1965)}, analyzes the fluctuations
of a infinitely thin and planar liquid-vapor interface
parallel to the $xy$-plane $\mathbb{R}^{2}$.  Local
deviations from the planar surface configuration are
represented by a random field $u(\mathbf{R})$, with
$\mathbf{R}=(x,y)\in\mathbb{R}^{2}$, which is considered
to be statistically independent of the particle distribution.
Consequently, the height-height correlation function
$\langle u(\mathbf{R})u(\mathbf{R}')\rangle_{\mathsf{CWT}}$
depends parametrically on the surface tension $\gamma$
but does not include any particle interactions explicitly.

In 1999, Mecke and Dietrich suggested a combination
of DFT and CWT. They use an isodensity criterion, i.e.,
$\rho\big(\mathbf{R},z=u(\mathbf{R})\big)=const.$,
in order to define the interface $u(\mathbf{R})$ implicitly
\cite{Mecke:6766(1999)}.  By means of a grand canonical
density functional $\Omega[\rho]$, they define an
effective interface Hamiltonian $\mathcal{H}[u]:=\Omega[\rho_{u}]-\Omega[\rho_{\circ}]$
as the difference between two free energies, associated
to two \emph{equilibrium} surface states. $\Omega[\rho_{\circ}]$
is the free energy of a planar interface parallel to
the $xy$-plane, while $\Omega[\rho_{u}]$ is associated
to a fixed but spatially varying surface $u(\mathbf{R})$.
An expansion of $\Omega[\rho_{u}]$ near $\Omega[\rho_{\circ}]$
(or $\rho_{u}(\mathbf{r})$ at $\rho_{\circ}(\mathbf{r})$,
respectively) allows for an explicit expression of
$\mathcal{H}[u]$ in terms of $u(\mathbf{R})$. By
taking curvature corrections of the fluctuating interface
into account, the resulting height correlation function
is governed by a wavelength dependent surface tension
$\gmd$. While the included van der Waals attractions
induce a decrease of $\gmd$, its increase at large
values of $q$ is considered as an indication for a
bending rigidity for liquid surfaces. $\gmd$ has been
confirmed in x-ray studies for several liquids with
different accuracy \cite{Fradin:871(2000), Mora:216101(2003), Li:136102(2004), Luo:4527(2006), Li:235426(2006), Lin:106(2005)}.

Although the combination of DFT and CWT has been generalized
to binary mixtures \cite{Hiester:184701(2006)}, the
results depend on the particular choice of the density
functional $\Omega[\rho]$ and on the validity of the
curvature corrections. There exist also difficulties
in analyzing the experimental scattering data in order
to obtain $\gmd$ \cite{Shpyrko:245423(2004)}. In
addition, recent numerical studies even cast doubt
on $\gmd$ \cite{Vink:134905(2005), Tarazona:196101(2007)}.
In Ref.\,\cite{Tarazona:196101(2007)} the authors
argue that $\gmd$ must decrease for large values of
$q$ if derived from a DFT as in Ref.\,\cite{Mecke:6766(1999)}.
Thus, alternative theoretical methods are mandatory
in order to elucidate the relation between density
correlations and interface correlations.

Here the height-height correlation function for a fluctuating
interface between two coexisting phases is derived
and expressed by means of general equilibrium properties
of the corresponding density-density correlations.
In principle, this approach is neither restricted to
planar systems nor to liquid-vapor interfaces. An explicit
expression for a wavelength-dependent surface tension
$\gamma(\mathbf{q})$ can be identified, which depends
on the direct correlation function $c(\mathbf{r},\mathbf{r}')$
(which is defined via the non-ideal gas contributions
of the \emph{inverse} density covariance function),
the equilibrium density profile $\rho_{\circ}(\mathbf{r})$
and, an operator relating density and surface configurations.
The expression for $\gamma(\mathbf{q})$ generalizes
in particular $\gmd$ and its derivation does not rely
on a free energy expansion, which is one of the main
criticism of Ref.\,\cite{Tarazona:196101(2007)}.

We consider an equilibrium state within the grand canonical
ensemble. Ensemble averages are denoted as $\langle...\rangle_{\circ}$.
Using the local microscopic density $\rho(\mathbf{r}):=\sum_{i=1}^{N}\delta(\mathbf{r}-\mathbf{r}_{i})$
we write $\rho_{\circ}(\mathbf{r}):=\langle\rho(\mathbf{r})\rangle_{\circ}$
for the spatially dependent one-particle equilibrium
density. We assume, that the system exhibits a single
interface between two coexisting phases. Its mathematical
specification is given below. Density fluctuations
are characterized by the density covariance function
$G(\mathbf{r},\mathbf{r}'):=\big\langle\,\big(\rho(\mathbf{r})-\rho_{\circ}(\mathbf{r})\big)\,\big(\rho(\mathbf{r}')-\rho_{\circ}(\mathbf{r}')\big)\,\big\rangle_{\circ}$.
Since fluctuations of the interface are not independent
of density fluctuations, the height-height correlation
function must be related to $G(\mathbf{r},\mathbf{r}')$.
In the following, we study this relationship.

It follows from probability theory, that a random field
$\rd(\mathbf{r})\in\mathbb{R}$ exists, which exhibits
the same mean value and correlations as $\rho(\mathbf{r})$,
i.e., $\overline{\rd(\mathbf{r})}=\rho_{\circ}(\mathbf{r})$
and $\overline{\big[\rd(\mathbf{r})-\rho_{\circ}(\mathbf{r})\big]\big[\rd(\mathbf{r}')-\rho_{\circ}(\mathbf{r}')\big]}=G(\mathbf{r},\mathbf{r}')$,
where the average $\overline{\,\stackrel{}{\ldots}\,}$
is taken with respect to a suitable probability measure
for $\rd$ \cite{Lifshits:1995}. In general, the system
under consideration is inhomogeneous due to the presence
of an interface and thus $G(\mathbf{r},\mathbf{r}')$
does \emph{not} depend only on $\mathbf{r}-\mathbf{r}'$.
Since the first and the second moment of $\rd(\mathbf{r})$
are given by $\rho_{\circ}(\mathbf{r})$ and the $G(\mathbf{r},\mathbf{r}')$,
$X(\mathbf{r})$ can be considered as, e.g., an inhomogeneous
Gaussian process. By construction, the local microscopic
density $\rho(\mathbf{r})$ and the random field $\rd(\mathbf{r})$
can be distinguished only by measurements or calculations
of their higher correlation functions. In particular,
the structure factor is not sufficient to determine
the difference between $\rd(\mathbf{r})$ and $\rho(\mathbf{r})$.
Thus, $\rd(\mathbf{r})$ is \emph{not} an equilibrium
density but mimics the first and the second equilibrium
moment of $\rho(\mathbf{r})$. We call $\rd(\mathbf{r})$
a random density. If, in addition, $\rd(\mathbf{r})$
is generated by another random field which mimics the
interface configurations, the correlations of this
underlying process are related to $G(\mathbf{r},\mathbf{r}')$.
In the following, this procedure and the relation between
the different correlation functions is discussed. 

Although we will focus later on planar interfaces,
we use first a slightly more abstract notation in order
to isolate the core definitions and ideas from additional
assumption about the system like its symmetry. 

\begin{figure}
\includegraphics[width=1\columnwidth,keepaspectratio]{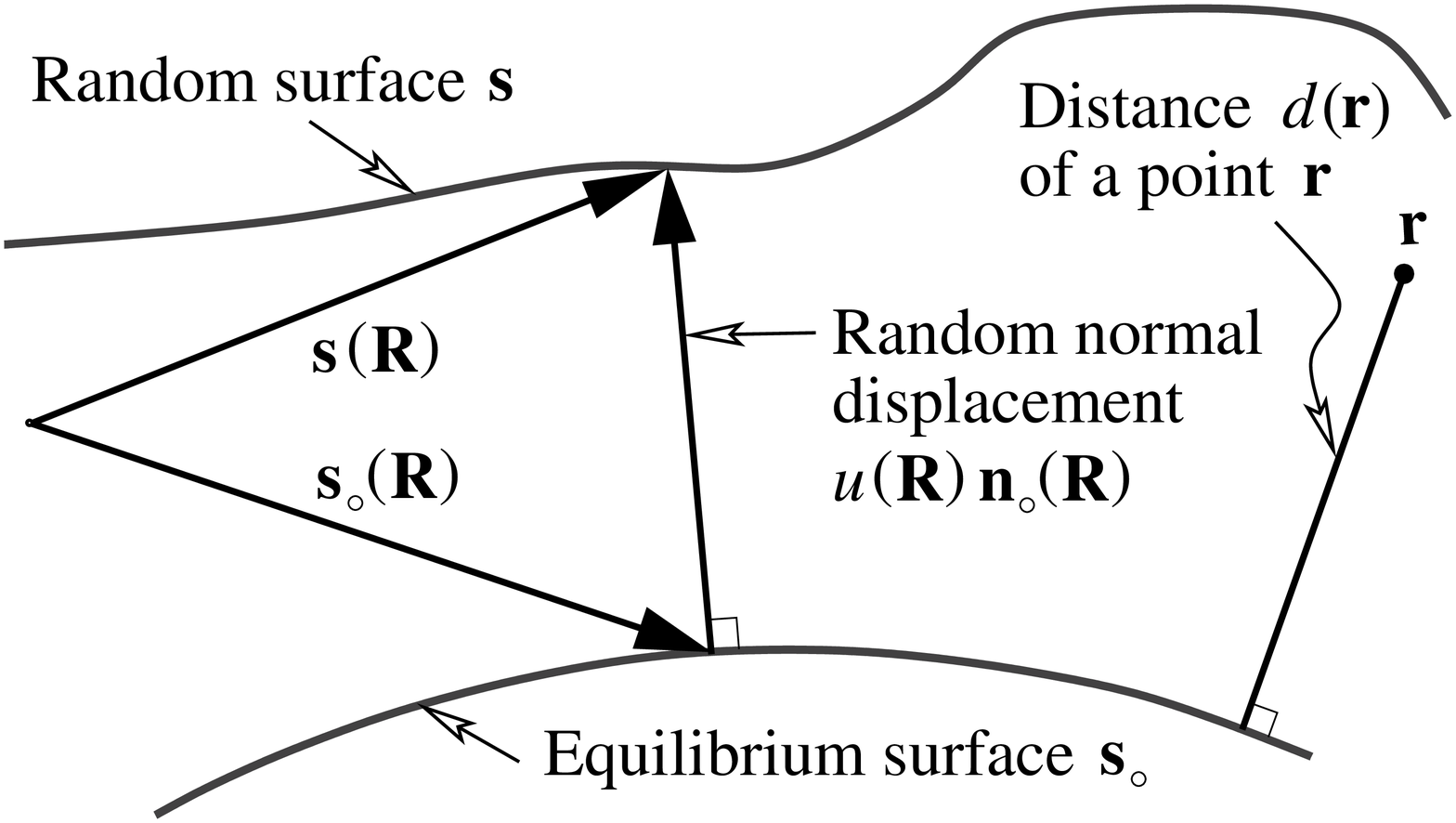}

\caption{\label{fig:normal_fluctuations}The random surface
$\rs$ is generated by random displacements $u(\mathbf{R})$
of $\rs_{\circ}(\mathbf{R})$ along its normal vector
$\mathbf{n}_{\circ}(\mathbf{R})$ (see Eq.\,\eqref{eq:fluctuating_distance}).
For simplicity, the surface patch $\rs_{\circ}$ is
drawn spherically. For each spatial point $\mathbf{r}$
we write $d\equiv d(\mathbf{r})$ for its (normal)
distance to the equilibrium interface $\rs_{\circ}$
(see Eq.\,\eqref{eq:r_in_n_o}). Random deviations
of the equilibrium density $\rho_{\circ}(\mathbf{r})$
are caused by the random displacements $u(\mathbf{R})$
of the interface as well as by a local change in volume
(see Eqs.\,\eqref{eq:rd_simple} and \eqref{eq:random_density_to_surface}).}
\end{figure}

Corresponding to the equilibrium density $\rho_{\circ}(\mathbf{r})$
we introduce an equilibrium phase boundary $\rs_{\circ}(\mathbf{R})\in\mathbb{R}^{3}$,
which depends on two parameters $\mathbf{R}\in M\subseteq\mathbb{R}^{2}$
for a suitable subset $M$ of $\mathbb{R}^{2}$. The
interface $\rs_{\circ}(\mathbf{R})$ is considered
to be an isodensity contour of $\rho_{\circ}(\mathbf{r})$,
i.e., $\rho_{\circ}\big(\mathbf{s}_{\circ}(\mathbf{R})\big)=\rho_{c}=const.$
for all $\mathbf{R}\in M$. Introducing the random
field $u(\mathbf{R})\in\mathbb{R}$, a fluctuating
interface $\rs(\mathbf{R})\in\mathbb{R}^{3}$ is considered
to be a random field which results from $\rs_{\circ}(\mathbf{R})$
by local normal displacements (see Fig.\,\ref{fig:normal_fluctuations}),
\begin{equation}
\rs(\mathbf{R})=\rs_{\circ}(\mathbf{R})+u(\mathbf{R})\,\mathbf{n}_{\circ}(\mathbf{R})\;,\label{eq:fluctuating_distance}\end{equation}
where $\mathbf{n}_{\circ}(\mathbf{R})$ is the unit
normal vector of the surface $\rs_{\circ}(\mathbf{R})$.
For instance, we expect $\mathbf{n}_{\circ}(x,y)=\mathbf{e}_{z}$
for a liquid-vapor interface in a homogeneous field
acting along the $z$-axis $\mathbf{e}_{z}$, while
for a spherical droplet it is $\mathbf{n}_{\circ}(\vartheta,\varphi)=\mathbf{e}_{r}(\vartheta,\varphi)$,
where $\mathbf{e}_{r}$ the radial unit vector. 

A relation between the random density $\rd(\mathbf{r})$
and the random surface $\rs(\mathbf{R})$ establishes
a relation between the density correlations $G(\mathbf{r},\mathbf{r}')$
and the correlations of $u(\mathbf{R})$. For this
purpose, we require the following conditions:\begin{subequations}\begin{eqnarray}
\rd\big(\rs(\mathbf{R})\big) & = & \rho_{\circ}\big(\rs_{\circ}(\mathbf{R})\big)=\rho_{c}\label{eq:1}\\
\rs\equiv\rs_{\circ} & \Rightarrow & \rd(\mathbf{r})=\rho_{\circ}(\mathbf{r})\mbox{ for all }\mathbf{r}\;,\label{eq:2}\end{eqnarray}
\label{eq:conditions}\end{subequations} where $\rs\equiv\rs_{\circ}$
means the identity for all $\mathbf{R}\in M$. The
condition Eq.\,\eqref{eq:1} generalizes the concept
of an isodensity contour to $\rd(\mathbf{r})$. The
condition \eqref{eq:2} states that $\rd(\mathbf{r})$
assumes its mean value if $\rs$ is identical to $\rs_{\circ}$.
Thus, we consider those $\rd(\mathbf{r})$ which result
from $\rho_{\circ}(\mathbf{r})$ due to a change in
the isodensity surface.

The above-mentioned conditions are not sufficient to
determine a unique relation between $\rd(\mathbf{r})$
and $\rs(\mathbf{R})$. But here we show in which way
different realizations of Eqs.\,(\ref{eq:conditions})
lead to different height correlations for $u(\mathbf{R})$
exploiting only the above-mentioned condition that
the auto-correlation of $\rd(\mathbf{r})-\rho_{\circ}(\mathbf{r})$
is given by $G(\mathbf{r},\mathbf{r}')$.%

We assume, that each point $\mathbf{r}$ can be represented
unambiguously within the normal coordinate system of
the equilibrium surface $\rs_{\circ}$, i.e., there
exist $d(\mathbf{r})\equiv d\in\mathbb{R}$ and $\mathbf{R}(\mathbf{r})\equiv\mathbf{R}\in M$
so that (see Fig.\,\ref{fig:normal_fluctuations})
\cite{Schroeder:551(2003)} \begin{equation}
\mathbf{r}=\rs_{\circ}(\mathbf{R})+d\,\mathbf{n}_{\circ}(\mathbf{R})\;.\label{eq:r_in_n_o}\end{equation}
A simple choice for $\rd(\mathbf{r})$ which fulfills
Eqs.\,(\ref{eq:conditions}) is \begin{eqnarray}
\rd(\mathbf{r}) & = & \rho_{\circ}\big(\mathbf{r}+\rs_{\circ}(\mathbf{R})-\rs(\mathbf{R})\big)\;,\label{eq:rd_simple}\end{eqnarray}
where $\mathbf{R}\equiv\mathbf{R}(\mathbf{r})$ (see
Eq.\,\eqref{eq:r_in_n_o}) \cite{remark}. By Eq.\,\eqref{eq:fluctuating_distance},
$\rd(\mathbf{r})$ is generated by random displacements
of the equilibrium density $\rho_{\circ}(\mathbf{r})$
along the normal vector $\mathbf{n}_{\circ}$ in whose
direction $\mathbf{r}$ lies. It is worth to note,
that Eq.\,\eqref{eq:rd_simple} is \emph{not} a result
but rather a choice for the parametrization of $\rd(\mathbf{r})$
based on Eqs.\,(\ref{eq:conditions}). Although Eq.\,\eqref{eq:rd_simple}
appears as the standard approach in order to describe
capillary wave fluctuations (of planar interfaces),
the specification of the conditions Eqs.\,(\ref{eq:conditions})
is an attempt to justify this approach on a more profound
principle. In particular, it does not rely on the thermodynamic
conditions as, e.g., the Gibbs dividing surface concept.
Here, we are interested the generalization of Eq.\,\eqref{eq:rd_simple}
which are consistent with that principle.

In general, a change in density at point $\mathbf{r}$
is obtained due to a change of the infinitesimally
small volume which contains the point $\mathbf{r}$.
This change of the local volume at point $\mathbf{r}$
might occur from a change of the equilibrium interface
even if the corresponding point on $\rs_{\circ}$ is
not shifted along $\mathbf{n}_{\circ}$. For instance,
a change from a spherical droplet to an ellipsoid changes
the local volume at a point $\mathbf{r}$ without moving
the corresponding point on $\rs_{\circ}$. Thus, a
change in curvature is one of these mechanisms. Obviously,
such changes in local volume depends on the distance
$d$ between $\mathbf{r}$ and $\rs_{\circ}$ (see
Eq.\,\eqref{eq:r_in_n_o} and Fig.\,\ref{fig:normal_fluctuations}).
Due to the isodensity condition this effect vanishes
for $d=u(\mathbf{R})$, i.e., if $\mathbf{r}$ lies
on $\rs(\mathbf{R})$. Equally, if the amplitude of
the interface fluctuations are sufficiently small,
the local volume within the bulk phases does not change.
Thus, this effect vanishes for $d\gg\xi$, where $\xi$
is the effective width of the interface, i.e., the
width of the transition region in which $\rho_{\circ}(\mathbf{r})$
differs considerably from the bulk densities. For step-like
profiles it is $\xi=0$. In such cases we do not expect
such a change in local volume due to other mechanisms.
A change in local volume that affects the local density
$\rho_{\circ}(\mathbf{r})$ can be associated with
a local compressibility. Therefore, the following generalizations
of Eq\@.\,(\ref{eq:rd_simple}) are intended to take
into account the local compressibility. 

The incompleteness of the standard capillary wave ansatz
Eq.\,\eqref{eq:rd_simple} can be understood equally
from the physical picture, that density fluctuations
are present in each spatial direction and not only
along the normal direction of the interface. This lack
in description could be adjusted by taking tangential
fluctuations (parallel to the interface) into account.
On the other hand, tangential density fluctuations
affect the normal density fluctuations due to the isodensity
condition of the interface. Therefore we expect an
additional term in Eq.\,\eqref{eq:rd_simple} which
projects tangential density fluctuation on normal density
fluctuations, similar as it is done in the projector
operator formalism. This projector includes the ratio
of correlation functions associated with tangential
and normal fluctuations, respectively. Furthermore,
it depends in general on the distance to the interface
and on the wavelength of the fluctuation: For large
distances the tangential density fluctuations are not
expected to affect the normal density fluctuation which
implies that the projected contribution disappears
independently of the wavelength of the fluctuation.%
{} Close to the interface, a long-wavelength tangential
density fluctuation will barely have an influence on
the local interfacial structure due to the isodensity
condition, i.e., such tangential fluctuations shall
be rather aligned by the global structure of the interface.
Thus, the projector is expected to vanish in that case.%
{} For the same reason, the local interfacial structure
depends on short-wavelength tangential fluctuation
close to the interface. For instance, the interface
might be bend locally due to a tangential density fluctuation
above or below the interface. At this point the interpretation
in terms of tangential density fluctuations becomes
similar to the above-mentioned local compressibility
picture.%
{} Therefore, the interfacial width $\xi$ can be considered
also as the range of the coupling between these two
types of fluctuations. 

After these pictorial remarks we introduce the $\mathbf{r}$-dependent
linear operator $\lop(\mathbf{r})$ which acts on the
surface $\rs(\mathbf{R})$ (and $\rs_{\circ}(\mathbf{R})$)
via\begin{equation}
\lop(\mathbf{r})\rs(\mathbf{R}):=\int_{M}\,\lopd(\mathbf{r};\mathbf{R},\mathbf{R}')\,\mathbf{s}(\mathbf{R}')\,\mbox{d}\mathbf{R}'\;,\label{eq:Linear_operator_3d}\end{equation}
with a $3\times3$-matrix $\lopd(\mathbf{r};\mathbf{R},\mathbf{R}')$
as integral kernel. Since $\lop(\mathbf{r})$ is intended
to model the effect of local compressibility we expect
$\lop(\mathbf{r})\rs_{\circ}(\mathbf{R})=\mathbf{0}$
for all $\mathbf{R}\in M$ and all $\mathbf{r}$ due
to the isodensity condition. Thus, it is $\lop(\mathbf{r})\rs(\mathbf{R})=\lop(\mathbf{r})\,\big(u(\mathbf{R})\mathbf{n}_{\circ}(\mathbf{R})\big)$.
Equally we assume $\lop(\mathbf{0})=\mathbf{0}$. For
this reason, a more general choice for $\rd(\mathbf{r})$
 which fulfills the conditions in Eqs.\,(\ref{eq:conditions})
is\begin{equation}
\rd(\mathbf{r})=\rho_{\circ}\big(\mathbf{r}-u(\mathbf{R})\mathbf{n}_{\circ}(\mathbf{R})+\lop(\mathbf{r}-\rs(\mathbf{R}))\rs(\mathbf{R})\big)\,,\label{eq:random_density_to_surface}\end{equation}
where $\mathbf{R}\equiv\mathbf{R}(\mathbf{r})$ (see
Eq.\,\eqref{eq:r_in_n_o}). As mentioned above,
this study elucidates the general influence of $\lop(\mathbf{r})$
on the height correlations. We will not derive its
particular form.%
{} From the physical meaning discussed above in terms
of tangential density fluctuations, it becomes clear,
that the integral kernel $\lopd(\mathbf{r};\mathbf{R},\mathbf{R}')$
should reflect the presence of an interfacial symmetry
of the equilibrium interface $\rs_{\circ}$. We will
demonstrate this principle below in the case of planar
interfaces.

An expansion of $\rd(\mathbf{r})$ given by Eq.\,\eqref{eq:random_density_to_surface}
with respect to $u(\mathbf{R})$ up to linear order
gives\begin{equation}
\rd(\mathbf{r})-\rho_{\circ}(\mathbf{r})\approx-\,\nabla\rho_{\circ}(\mathbf{r})\,\big{[}\mathbf{1}_{3}-\lop(d\mathbf{n}_{\circ}(\mathbf{R}))\,\big]\, u(\mathbf{R})\mathbf{n}_{\circ}(\mathbf{R})\,,\label{eq:rd_linear_expansion}\end{equation}
where $\mathbf{1}_{3}$ means the $3\times3$ unit
matrix, $d\equiv d(\mathbf{r})$ is the distance of
$\mathbf{r}$ to the surface $\rs_{\circ}$ and $\mathbf{R}\equiv\mathbf{R}(\mathbf{r})$
(see Eq.\,\eqref{eq:r_in_n_o} and Fig.\,\ref{fig:normal_fluctuations}).
$\lop(d\mathbf{n}_{\circ}(\mathbf{R}))u(\mathbf{R})\mathbf{n}_{\circ}(\mathbf{R})$
can be decomposed into normal and tangential contributions
with respect to $\rs_{\circ}(\mathbf{R})$ at each
$\mathbf{R}$ which allows for a general derivation
of the height correlations. Here, we demonstrate the
main idea for \emph{planar} interfaces. More general
cases like the spherical or the cylindrical interface
differ by their coordinate system which makes primarily
the notation more complex. In addition, closed interfaces
(e.g. droplets) are subject to an additional volume
constraint for stability reasons; these exclude certain
types of fluctuations. The crucial point in all cases
is the symmetry of the equilibrium interface which
is reflected by the correlation function $G(\mathbf{r},\mathbf{r}')$
and the integral kernel $\lopd(\mathbf{r};\mathbf{R},\mathbf{R}')$. 

In the following discussion of height correlations
of infinite planar interfaces ($\mathbf{R}\equiv(R_{1},R_{2})\in M=\mathbb{R}^{2}$),
the equilibrium surface $\rs_{\circ}(\mathbf{R})=R_{1}\,\mathbf{e}_{x}+R_{2}\,\mathbf{e}_{y}+z_{\circ}\,\mathbf{e}_{z}$,
with a constant $z_{\circ}$, lies parallel to $(x,y)$-plane
so that $\mathbf{n}_{\circ}(\mathbf{R})=\mathbf{e}_{z}$.
The equilibrium density $\rho_{\circ}(\mathbf{r})\equiv\rho_{\circ}(z)$
depends only on the normal distance $z=z_{\circ}+d(\mathbf{r})$
to the surface $\rs_{\circ}$ (see Eq.\,\eqref{eq:r_in_n_o}
and Fig.\,\ref{fig:normal_fluctuations}). Consequently,
the density correlation function $G(\mathbf{r},\mathbf{r}')\equiv G(z,z',\mathbf{R}-\mathbf{R}')$
is homogeneous with respect to the lateral coordinates.
Equally, the integral kernel possesses the same homogeneity,
i.e., $\lopd(d\mathbf{n}_{\circ};\mathbf{R},\mathbf{R}')\equiv\lopd(z-c;\mathbf{R}-\mathbf{R}')$.
Since $\rho_{\circ}(z)$ depends only on $z$, it is
$\partial_{x}\rho_{\circ}=\partial_{y}\rho_{\circ}=0$
and Eq.\,\eqref{eq:rd_linear_expansion} becomes\begin{equation}
\rd(z,\mathbf{R})-\rho_{\circ}(z)\approx-\,\partial_{z}\rho_{\circ}(z)\,[1-\lop(z-z_{\circ})]\, u(\mathbf{R})\;,\label{eq:rd_linear_expansion_planar}\end{equation}
where $\lop(z-z_{\circ})u(\mathbf{R})=\int_{\mathbb{R}^{2}}L(z-z_{\circ};\mathbf{R}-\mathbf{R}')\, u(\mathbf{R}')\,\mbox{d}\mathbf{R}'$
with $\lop(0)=0$ is a linear operator which depends
on $z-z_{\circ}$ and which acts on the random variable
$u(\mathbf{R})$ (see also Eq.\,\eqref{eq:Linear_operator_3d}).
The correlation function for $\rd(z,\mathbf{R})-\rho_{\circ}(z)$
is given by $G(z,z',\mathbf{R}-\mathbf{R}')$. By taking
the Fourier-transformation with respect to the lateral
coordinates, we obtain from Eq.\,\eqref{eq:rd_linear_expansion_planar}\begin{eqnarray}
 &  & 2\pi\,\delta(\mathbf{q}+\mathbf{q}')\,\hat{G}(z,z',\mathbf{q})\nonumber \\
 & = & \partial_{z}\rho_{\circ}(z)\partial_{z}\rho_{\circ}(z')\times\label{eq:G_planar_interface_Fourier}\\
 &  & \qquad[1-\hat{\lopds}(z-z_{\circ};\mathbf{q})][1-\hat{\lopds}(z'-z_{\circ};\mathbf{q}')]\,\overline{\,\hat{u}(\mathbf{q})\,\hat{u}(\mathbf{q}')\,}.\nonumber \end{eqnarray}
The interpretation of $\lop$ as a projector of tangential
density fluctuations on normal density fluctuations
implies, that $\hat{\lopds}$ includes the ratio of
Fourier components of the transverse auto-correlations
and the height auto-correlation $\overline{\,\hat{u}(\mathbf{q})\hat{u}(\mathbf{q}')\,}$.
Therefore and from the discussion above it becomes
clear, that, in particular for larger $q$ values,
$\hat{G}$ is reasonably governed by the transverse
correlations and not only by the height correlations
as in the case $\lop\equiv0$. 

$\hat{G}(z,z',\mathbf{q})$ is a positive definite
function and symmetric in $z$ and $z'$. This allows
for an expansion of Eq.\,\eqref{eq:G_planar_interface_Fourier}
in terms of eigenfunctions of $\hat{G}(z,z',\mathbf{q})$,
which leads to a generalized version of Wertheim's
eigenfunction analysis of the correlations in a planar
liquid-gas interface for small values of $|\mathbf{q}|$
\cite{Wertheim:2377(1976)}. The need of such a generalized
approach has been already concluded from numerical
investigations of the eigenfunction ansatz \cite{Stecki:7967(1997)}.
Here, we first ask in which way the term $\hat{L}(z-z_{\circ};\mathbf{q})$
affects the \emph{general form} of $\overline{\,\hat{u}(\mathbf{q})\hat{u}(\mathbf{q}')\,}$
which will be answered below \emph{without} using an
expansion in eigenfunctions. For low $q$ values, the
result becomes equal to the expression from Wertheim's
approach.%
{} It is important to bear in mind, that $\hat{G}(z,z',\mathbf{q})$
results from a thermodynamic average procedure while
the height correlation function $\overline{\,\hat{u}(\mathbf{q})\hat{u}(\mathbf{q}')\,}$
stems from an unknown probability measure. Therefore,
Eq.\,\eqref{eq:G_planar_interface_Fourier} is in
particular a manifestation of the required thermodynamic
consistency condition mentioned above. %
{}%
{}

The \emph{inverse} density correlation function $G^{-1}(\mathbf{r},\mathbf{r}')$
is defined by $\int\mbox{d}\mathbf{r}'\, G(\mathbf{r},\mathbf{r}')G^{-1}(\mathbf{r}',\mathbf{r}'')=\delta(\mathbf{r}-\mathbf{r}'')$.
This implies from Eq.\,\eqref{eq:G_planar_interface_Fourier}
for the height correlation function\begin{subequations}\begin{eqnarray}
\overline{\,\hat{u}(\mathbf{q})\hat{u}(\mathbf{q}')\,} & = & 2\pi\,\delta(\mathbf{q}+\mathbf{q}')\, C(\mathbf{q})\label{eq:inf_height_general_1}\\
C(\mathbf{q}) & := & \Big[\,\iint_{-\infty}^{\infty}\mbox{d}z\mbox{d}z'\,\partial_{z}\rho_{\circ}(z)\partial_{z'}\rho_{\circ}(z')\,\times\label{eq:inf_height_general_2}\\
 &  & \!\!\!\!\!\!\!\!\!\!\!\!\!\!\!\!\!\!\!\!\!\!\!\!\!\!\!\!\!\!\!\!\!\!\!\!\!\!\!\![1-\hat{\lopds}(z-z_{\circ};-\mathbf{q})][1-\hat{\lopds}(z'-z_{\circ};\mathbf{q})]\,\hat{G}^{-1}(z,z',\mathbf{q})\,\Big]^{-1}\;.\nonumber \end{eqnarray}
\label{eq:inv_height_general}\end{subequations}$G^{-1}$
is typically written as $G^{-1}(\mathbf{r},\mathbf{r}')=\delta(\mathbf{r-}\mathbf{r}')/\rho_{\circ}(\mathbf{r})-c(\mathbf{r},\mathbf{r}')$,
where $c(\mathbf{r},\mathbf{r}')$ is called direct
correlation function \cite{Evans:143(1979)}. In Fourier
space, we have for the planar interfaces $\hat{G}^{-1}(z,z',\mathbf{q})=\delta(z-z')/\rho_{\circ}(z)-\hat{c}(z,z',\mathbf{q})$.
Furthermore, the equilibrium density $\rho_{\circ}(z)$
fulfills the generalized barometric law, i.e., the
equation $\ln\lambda_{\mathsf{th}}^{3}\rho_{\circ}(z)-c^{(1)}(z)+\beta V^{\mathsf{ext}}(z)=\beta\mu$,
where $\lambda_{\mathsf{th}}$ is the thermal de-Broglie
wavelength, $c^{(1)}(z)$ means the effective one-particle
potential, $V^{\mathsf{ext}}(z)$ is an external potential
and, $\mu$ means the chemical potential \cite{Rowlinson:1982}.
By taking the derivative of the equilibrium condition
for $\rho_{\circ}(z)$ and bearing in mind the relation
$\nabla c^{(1)}(\mathbf{r})=\int\mbox{d}\mathbf{r}'\, c(\mathbf{r},\mathbf{r}')\nabla'\rho_{\circ}(\mathbf{r}')$
(see Ref.~\cite{Rowlinson:1982}) we can rewrite the
ideal gas contribution in $G^{-1}$. This gives\begin{align}
\hat{G}^{-1}(z,z',\mathbf{q}) & =-\beta\,\frac{\delta(z-z')\,\partial_{z}V^{\mathsf{ext}}(z)}{\partial_{z}\rho_{\circ}(z)}-\hat{c}(z,z',\mathbf{q})\label{eq:invG_o_explicit}\\
 & +\frac{\delta(z-z')}{\partial_{z}\rho_{\circ}(z)}\,\int\mbox{d}z''\,\hat{c}(z,z'',\mathbf{0})\partial_{z}\rho_{\circ}(z'')\;.\nonumber \end{align}
By combining Eqs.\,(\ref{eq:inv_height_general}) and
\eqref{eq:invG_o_explicit} we obtain%
{}\begin{subequations}\begin{eqnarray}
\beta\,\overline{\,\hat{u}(\mathbf{q})\,\hat{u}(\mathbf{q}')\,} & = & \frac{2\pi\,\delta(\mathbf{q}+\mathbf{q}')}{v^{\mathsf{ext}}(\mathbf{q})+\eta(\mathbf{q})}\;,\label{eq:Inverse-height-height_correlation_planar}\end{eqnarray}
where\begin{equation}
v^{\mathsf{ext}}(\mathbf{q}):=-\int_{-\infty}^{\infty}\mbox{d}z\,\partial_{z}V^{\mathsf{ext}}(z)\partial_{z}\rho_{\circ}(z)\,|1-\hat{L}(z-z_{\circ};\mathbf{q})|^{2}\label{eq:Def_beta_v_ext}\end{equation}
 and\begin{align}
\eta(\mathbf{q}) & :=\frac{1}{\beta}\,\iint_{-\infty}^{\infty}\mbox{d}z\mbox{d}z'\,\partial_{z}\rho_{\circ}(z)\partial_{z'}\rho_{\circ}(z')\,\times\label{eq:Def_eta_planar}\\
 & \qquad\Big[\,\big{(}\hat{c}(z,z',\mathbf{0})-\,\hat{c}(z,z',\mathbf{q})\big{)}\,\big|1-\hat{\lopds}(z-z_{\circ};\mathbf{q})\big|^{2}\nonumber \\
 & \qquad+\frac{1}{2}\,\hat{c}(z,z',\mathbf{q})\,\big{|}\hat{\lopds}(z-z_{\circ};\mathbf{q})\,-\hat{\lopds}(z'-z_{\circ};\mathbf{q})\big{|}^{2}\,\Big]\;.\nonumber \end{align}
\label{eq:inv_height_corr}\end{subequations}By construction,
the  height-height correlations given by Eqs.\,(\ref{eq:inv_height_corr})
are thermodynamically consistent with the density covariance
function $G(\mathbf{r},\mathbf{r}')$. From the derivation
above the $\mathbf{q}$-dependence of $\eta(\mathbf{q})$
comes in via the direct correlation function $c(\mathbf{r},\mathbf{r}')$
and the kernel $\hat{\lopds}(z;\mathbf{q})$ of the
linear operator $\lop(z)$. While $c(\mathbf{r},\mathbf{r}')$
represents the particle interactions, $\lop(z)$ takes
into account the change of the equilibrium density
$\rho_{\circ}(z)$ due to a local volume change at
point $z$, i.e., the local compressibility, or the
ratio of tangential density correlations and normal
density correlations, respectively.

The derivation of the explicit expression for $\overline{\,\hat{u}(\mathbf{q})\,\hat{u}(\mathbf{q}')\,}$,
i.e., from Eq.\,\eqref{eq:G_planar_interface_Fourier}
to Eqs.\,(\ref{eq:inv_height_corr}), is based on
the definition for $\hat{G}^{-1}(z,z',\mathbf{q})$
and the generalized barometric law. Therefore, \emph{any
choice} for $\hat{L}$ leads to a height correlation
function Eqs.\,(\ref{eq:inv_height_corr}) which is
consistent with the density covariance $G(\mathbf{r},\mathbf{r}')$.
Putting Eq.\,\eqref{eq:Inverse-height-height_correlation_planar}
back into Eq.\,\eqref{eq:G_planar_interface_Fourier}
it follows\begin{eqnarray}
\beta\hat{G}(z,z',\mathbf{q}) & = & \partial\rho_{\circ}(z)\partial\rho_{\circ}(z')\times\label{eq:G_planar_interface_Fourier_final}\\
 &  & \frac{[1-\hat{\lopds}(z-z_{\circ};\mathbf{q})][1-\hat{\lopds}(z'-z_{\circ};-\mathbf{q})]}{v^{\mathsf{ext}}(\mathbf{q})+\eta(\mathbf{q})}\;,\nonumber \end{eqnarray}
where the functional dependence of $v^{\mathsf{ext}}(\mathbf{q})$
and $\eta(\mathbf{q})$ on $\hat{\lopds}(z-z_{\circ};\mathbf{q})$
is given by Eq.\,\eqref{eq:Def_beta_v_ext} and \eqref{eq:Def_eta_planar},
respectively. Consequently, for a given $\hat{G}(z,z',\mathbf{q})$
the Eq.\,\eqref{eq:G_planar_interface_Fourier_final}
is a definition for $\hat{\lopds}(z-z_{\circ};\mathbf{q})$
(possibly limited by the linearization made in Eq.\,\eqref{eq:rd_linear_expansion_planar}).
For instance, using the total correlation function
$\hat{h}(z,z',\mathbf{R}-\mathbf{R}')$, it is $\hat{G}(z,z',\mathbf{q})=\delta(z-z')\rho_{\circ}(z)+\rho_{\circ}(z)\rho_{\circ}(z')\hat{h}(z,z',\mathbf{q})$.
From the Ornstein-Zernike relation for inhomogeneous
systems combined with some closure relation, an approximate
scheme for $\hat{h}(z,z',\mathbf{q})$ can be applied
in order to get a functional equation for $\hat{\lopds}(z-z_{\circ};\mathbf{q})$.%
{} In reverse, a given $\hat{\lopds}$ implies a particular
form of $\hat{G}(z,z',\mathbf{q})$, where the physical
meaning of $\hat{L}$ (as discussed above Eq.\,\eqref{eq:Linear_operator_3d})
might serve as a guidance. The corresponding height
correlation function is in any case consistently given
by the Eqs.\,(\ref{eq:inv_height_corr}). That means
in particular for larger $q$ values, that a sole investigation
of $\overline{\,\hat{u}(\mathbf{q})\,\hat{u}(\mathbf{q}')\,}$,
i.e., a sole examination of Eqs.\,(\ref{eq:inv_height_corr}),
is not sufficient in order to check the reliability
of a given $\hat{\lopds}(z-z_{\circ};\mathbf{q})$
\cite{linear_response_remark}.%
{}

The only approximations in the realization of Eqs.\,(\ref{eq:conditions})
are those in Eq.\,\eqref{eq:random_density_to_surface}
or \eqref{eq:rd_linear_expansion}, respectively. Surprisingly,
the concept of an effective interface Hamiltonian or
an expansion of free energies, respectively, is not
needed in order to derive the general structure of
height correlations which are consistent with the underlying
density correlations. Nevertheless, the Eqs.\,(\ref{eq:inv_height_corr})
and in particular Eq.\,\eqref{eq:Def_eta_planar}
can be compared to former expression for the height
correlations based on such principles. To do so, it
is convenient to consider $\gamma(q):=\eta(|\mathbf{q}|)/q^{2}$
which is referred to in literature as a wavelength
dependent surface energy density for isotropic interfaces.%
{}

The macroscopic surface tension $\gamma(0)$ has been
derived in Ref.\,\cite{Triezenberg:1183(1972)} and
follows from Eq.\,\eqref{eq:Def_eta_planar}, if $\hat{\lopds}(z;\mathbf{q})\rightarrow0$
for $q\rightarrow0$. For step-like profiles $\partial_{z}\rho_{\circ}(z)\sim\delta(z-z_{\circ})$
the  the $\hat{\lopds}$-terms drop out due to $\hat{\lopds}(0,\mathbf{q})=0$.
In that case, Eq.\,\eqref{eq:Def_eta_planar} reduces
to the result derived in Ref.\,\cite{Napiorkowski:1836(1993)}
if the same expression for direct correlation function
provided by the density functional in \cite{Napiorkowski:1836(1993)}
is used (see Eq.\,(4.6) in Ref.\,\cite{Napiorkowski:1836(1993)}).
Both cases reflect some properties of $\hat{L}$ which
we concluded from the general discussion above, i.e.,
the long-wavelength limit and the $\xi=0$ case.%
{}

The Eqs.\,(\ref{eq:inv_height_corr}) include also
the energy density for surface excitations derived
by Mecke and Dietrich in Gaussian approximation \cite{Mecke:6766(1999)}.
To see that, we first note, that the particular density
functional used in \cite{Mecke:6766(1999)} implies
an explicit expression $\hat{G}_{\mathsf{MD}}^{-1}(z,z',q)$
for the inverse density covariance function and thus
for the direct correlation function $\hat{c}_{\mathsf{MD}}(z,z',q)=\frac{\delta(z-z')}{\rho_{\circ}(z)}-\hat{G}_{\mathsf{MD}}^{-1}(z,z',q)$
\cite{remark2}. In Ref.\,\cite{Mecke:6766(1999)},
the external potential is the homogeneous gravity potential
$V_{\mathsf{MD}}^{\mathsf{ext}}(z)=mgz$ and the equilibrium
interface is located at $z_{\circ}=0$ (see Eq.\,(2.4)
in \cite{Mecke:6766(1999)}).%
{} By repeating the derivation of the Eqs.\,(\ref{eq:inv_height_corr})
from Eq.\,(\ref{eq:inv_height_general}), one applies
Eq.\,\eqref{eq:invG_o_explicit} with $V^{\mathsf{ext}}(z)\equiv V_{\mathsf{MD}}^{\mathsf{ext}}(z)$
and $\hat{c}(z,z',q)\equiv\hat{c}_{\mathsf{MD}}(z,z',q)$
except for the term $\partial_{z}\rho_{\circ}(z)\partial_{z'}\rho_{\circ}(z')\,\hat{\lopds}^{*}(z;q)\hat{\lopds}(z';q)\,\hat{G}^{-1}(z,z',q)$
which appears in Eq.\,(\ref{eq:inv_height_general}).
Therein, one uses $\hat{G}_{\mathsf{MD}}^{-1}(z,z',q)$
instead of Eq.\,\eqref{eq:invG_o_explicit} (which
leads to the $\kappa-\tilde{\kappa}_{0}^{(HH)}(q)$
contribution in Eq.\,(3.11) in \cite{Mecke:6766(1999)}).
By setting $\partial_{z}\rho_{\circ}(z)\,\hat{\lopds}(z;\mathbf{q})\equiv-\, q^{2}\rho_{H}(z)$
(where $\rho_{H}$ is given by the Eqs.\,(3.27) and
(3.31) in Ref.\,\cite{Mecke:6766(1999)} without further
derivation) the resulting expression for $\overline{\,\hat{u}(\mathbf{q})\,\hat{u}(\mathbf{q}')\,}$
and $\eta(q)/q^{2}$ are equal to Eqs.\,(4.1), (4.2)
and (3.11), respectively, in Ref.\,\cite{Mecke:6766(1999)}
(see also Eqs.\,(2.28) and (3.10) in \cite{Mecke:6766(1999)}).%
{} From the explicit expression for $\rho_{H}(z)$ and
$\rho_{\circ}(z)$ assumed in \cite{Mecke:6766(1999)}
we obtain $\hat{\lopds}_{\mathsf{MD}}(z,\mathbf{q})=\frac{C_{H}}{\pi}\,\xi\, z\,\sinh(\frac{z}{2\xi})\, q^{2}$
\cite{remark3}. Eq.\,\eqref{eq:G_planar_interface_Fourier_final}
implies, that $\hat{\lopds}_{\mathsf{MD}}(z,q)$ combined
with $\hat{c}_{\mathsf{MD}}(z,z',q)$ provide rather
a particular model for the density covariance function
$\hat{G}(z,z',q)$ than a model for the height correlations,
only. Therefore, in order to test the reliability of
the resulting expressions in computer simulations it
might be easier to check Eq.\,\eqref{eq:G_planar_interface_Fourier_final}
instead of Eqs.\,(\ref{eq:inv_height_corr}) via arguable
numerical procedures to define the position of the
fluctuating interface. 

The form of $\hat{\lopds}_{\mathsf{MD}}(z,\mathbf{q})$
implies, that the operator $\lop_{\mathsf{MD}}(z)$
is not bounded for $z\neq0$, since $\lop_{\mathsf{MD}}(z)\sim\nabla^{2}$
so that the kernel $\lopds_{\mathsf{MD}}(z,\mathbf{R}-\mathbf{R}')$
involves the $\delta$-distribution and its second
derivative in $x$- and $y$-direction. From Eq.\,\eqref{eq:random_density_to_surface}
follows that $\rd(\mathbf{r})\in[\rho_{\circ}^{+},\rho_{\circ}^{-}]$
where $\rho_{\circ}^{\pm}:=\rho_{\circ}(z\rightarrow\pm\infty)$
denote the bulk densities. Therefore, a realization
$u(\mathbf{R})$ of the interface with $u(\mathbf{R}_{p})=|\nabla u(\mathbf{R}_{p})|=0$
and $|\nabla^{2}u(\mathbf{R}_{p})|=\infty$ for a particular
point $\mathbf{R}_{p}$ induces $\rd(z\neq0,\mathbf{R}_{p})=\rho_{\circ}^{\pm}$
\cite{remark4}. On the other hand, the linearization
of Eq.\,\eqref{eq:random_density_to_surface} (Eq.\,\eqref{eq:rd_linear_expansion_planar}
for planar interfaces) with $\lop\equiv\lop_{\mathsf{MD}}$
allows for $\rd(\mathbf{r})\in[-\infty,\infty]$ even
if $u(\mathbf{R})$ is bounded, which seems not reasonable
from the physical point of view.  Therefore, as long
as one would like to linearize Eq.\,\eqref{eq:random_density_to_surface},
one has to limit the applicability of $\lop_{\mathsf{MD}}(z)$
to a particular set of interface configurations whose
second derivatives are also bounded or equivalently,
one has to limit the $q$-range of the kernel $\hat{\lopds}_{\mathsf{MD}}(z,\mathbf{q})$
\cite{remark5}. 

The limitation of the set of interface configurations
$\{ u(\mathbf{R})\}$ restricts the set of modeled
density configurations. If $\{\rd(\mathbf{r})\}$ denotes
the set of all density configurations, an optimal sampling
of $\{\rd(\mathbf{r})\}$ would capture the set of
all \emph{relevant} density configurations $\{\rd(\mathbf{r})\}_{\mathsf{rel}}\subseteq\{\rd(\mathbf{r})\}$
which, in our case, contribute to the density covariance
function $G(\mathbf{r},\mathbf{r}')$. The representation
of $\rd(\mathbf{r})$ by the interfacial field $u(\mathbf{R})$
leads to a set of density configurations $\{\rd(\mathbf{r})\}_{\mathsf{if}}$
which gives in general not the optimal sampling of
$\{\rd(\mathbf{r})\}$ such that $\{\rd(\mathbf{r})\}_{\mathsf{if}}\subset\{\rd(\mathbf{r})\}_{\mathsf{rel}}$.
Therefore one should allow for the \emph{largest} set
of interface configurations $\{ u(\mathbf{R})\}$ (whose
Fourier transform is continuous, for instance) in order
to exhaust $\{\rd(\mathbf{r})\}_{\mathsf{rel}}$ in
an optimal way, which corresponds to an approximation
of $G(\mathbf{r},\mathbf{r}')$ by the height correlation
function via Eq.\,\eqref{eq:G_planar_interface_Fourier_final}
in an optimal way. As an example, one may allow for
all interface configurations $u(\mathbf{R})$ with
$\Vert u\Vert_{1}:=\int_{\mathbb{R}^{2}}|u(\mathbf{R})|\,\mbox{d}\mathbf{R}<\infty$
and in order to use Eq.\,\eqref{eq:rd_linear_expansion_planar},
one may require $|\lopds(z,\mathbf{R})|<\infty$ for
all $\mathbf{R}\in\mathbb{R}^{2}$ so that $\lop(z)$
becomes a bounded operator for every $|z|<\infty$
\cite{remark6}. 

The operator $\lop_{\mathsf{MD}}(z\neq0)\sim\nabla^{2}$
is associated to the \emph{local} curvature of the
interface $u(\mathbf{R})$, since it results from a
truncated curvature expansion of the corresponding
density configuration (see Eq.\,(2.17) in \cite{Mecke:6766(1999)}).
Consequently, the coefficient of the resulting $q^{2}$-increase
in $\gmd$ has been interpreted as the bending rigidity
of the liquid interface. The above considerations about
the boundedness of $\lop(z)$ then suggest, that the
influnce of the local curvature of the interface and
the effect of bending rigidity can be probably not
continued to arbitrary small scales although the random
density $\rd(\mathbf{r})$ as well as the random interface
$u(\mathbf{R})$ \emph{are} defined on all length scales.
Nevertheless, since $\lop(z)u(\mathbf{R})$ is written
as a convolution integral (see Eq.\,\eqref{eq:rd_linear_expansion_planar}
below) and $\lop(z)$ is bounded, we thus may imagine
that $\lop(z)$ picks up the \emph{non-local} effects
of the interface $u(\mathbf{R})$ on the density configuration
$\rd(\mathbf{r})$. In a similar manner, the influence
of non-locality has been mentioned also in the framework
of short-ranged wetting \cite{Parry:wetting}. 

In our treatment of interface fluctuations, the distorted
interface $\rs(\mathbf{R})$ is considered to be generated
by local \emph{random} displacements $u(\mathbf{R})$
along the normal vector $\mathbf{n}_{\circ}(\mathbf{R})$
of the equilibrium interface $\rs_{\circ}(\mathbf{R})$
(see Eq.\,\eqref{eq:fluctuating_distance} and Fig.\,\eqref{fig:normal_fluctuations}).
By linking the random displacements to a change in
density via Eq.\,\eqref{eq:random_density_to_surface},
the density correlations $G(\mathbf{r},\mathbf{r}')$
govern the interface correlations $\overline{\, u(\mathbf{R})u(\mathbf{R}')\,}$.
The situation is in a sense \emph{reverse} to those
in stochastic differential equations, where the mean
value and the correlations of added noise terms (which
correspond to $u(\mathbf{R})$ here) are specified
in order to study its influence on a variable of interest
(which is $\rho(\mathbf{r})$ in our case). In general,
the noise contribution within a single realization
of the process can not be determined unambiguously.
Similarly, the difficulty to obtain $\overline{\, u(\mathbf{R})u(\mathbf{R}')\,}$
from numerical simulations arises from the problem
to identify the random interface, i.e., the realization
of the related (lower dimensional) stochastic process,
in a particle configuration \cite{remark,Chacon:166103(2003)}.
Consequently, any recipe that determines the random
interface in computer simulations works accurately
with a certain probability but can not be exact. As
pointed out in Ref.\,\cite{Tarazona:196101(2007)},
the precision of such procedures is enhanced the more
microscopic information of the system is included,
i.e., the more $n$-particle correlations are taken
into account to define the random interface numerically.

At this point we come back to the criticism of Tarazona,
Checa, and Chac\'{o}n \cite{Tarazona:196101(2007)}
concerning the Mecke/Dietrich approach. In order to
obtain the same expression for the height correlation
function as published in \cite{Mecke:6766(1999)} we
have neither used an effective interface Hamiltonian
nor a restricted variational principle as assumed in
\cite{Mecke:6766(1999)}. Formally, the random density
$\rd(\mathbf{r})$ corresponds to $\rho_{f}(\mathbf{r})$
in Ref.\,\cite{Mecke:6766(1999)} (see Eq.\,(2.6)
in \cite{Mecke:6766(1999)}) but the crucial difference
consists in the assumption, that $\rho_{f}(\mathbf{r})$
in \cite{Mecke:6766(1999)} minimizes the given density
functional with an additional isodensity condition
(see Eq.\,(2.5) in \cite{Mecke:6766(1999)}). While
the isodensity condition for $\rho_{f}$ in \cite{Mecke:6766(1999)}
is similar to the isodensity condition Eq.\,\eqref{eq:1}
for $\rd(\mathbf{r})$, no additional equation resulting
from a minimization procedure is required for $\rd(\mathbf{r})$,
i.e., the realization of Eqs.\,(\ref{eq:conditions})
are those in Eq.\,\eqref{eq:random_density_to_surface}
or \eqref{eq:rd_linear_expansion}, respectively. In
order to extract the height correlation function or
$\gmd$, respectively, from computer simulations, the
authors of Ref.\,\cite{Tarazona:196101(2007)} have
taken this minimization condition for $\rho_{f}$ in
\cite{Mecke:6766(1999)} seriously into account. Their
results differ considerably from the predicted $\gmd$
due to general features of density functionals as explained
in Ref.\,\cite{Tarazona:196101(2007)}. Since $\rho_{f}$
fulfills a minimization condition for a density functional,
it includes capillary waves on small wavelengths which
can not be separated as height fluctuations (see Fig.\,2
in \cite{Tarazona:196101(2007)}). While the numerical
analysis shows, that the minimization condition leads
to a different $\gmd$ than predicted, our approach
shows, that $\gmd$ can be derived indeed \emph{without}
an additional (minimization) condition. In other words,
the results in \cite{Tarazona:196101(2007)} do not
necessarily imply that $\gmd$ is structurally incorrect
because no density functional and related minimization
procedures are needed in order to derive $\gmd$. The
only dubious quantity that remains is $\hat{\lopds}(z-z_{\circ},q)$
or $\hat{\lopds}_{\mathsf{MD}}(z,q)$, respectively,
and from the derivation above it becomes clear, that
a model for $\hat{\lopds}(z-z_{\circ},q)$ is a different
task, that can not be solved within a capillary wave
theory. In particular, our derivation implies that
$\hat{\lopds}(z-z_{\circ},q)$ or $\hat{\lopds}_{\mathsf{MD}}(z,q)$
should be discussed rather in terms of Eq.\,\eqref{eq:G_planar_interface_Fourier_final}
than only in terms of Eqs.\,(\ref{eq:inv_height_corr}).
This suggestion if fully consistent with the conclusion
in \cite{Tarazona:196101(2007)}, that the capillary
wave problem can not be solved (numerically) on short
wavelengths by taking into account only the one particle
distribution.

Above we have discussed in detail the $q$-dependence
of $\hat{\lopds}(z,q)$ or $\hat{\lopds}_{\mathsf{MD}}(z,q)$,
respectively. Finally we briefly discuss a recently
published study on planar colloid-polymer interfaces
by Blokhuis, Kuipers and Vink \cite{Blokhuis:086101(2008)}
as an application of Eq.\,\eqref{eq:G_planar_interface_Fourier_final}
(with $z_{\circ}=0$) in order to show that also the
$z$-dependence of $\hat{\lopds}(z,q)$ plays a crucial
role. In Ref.\,\cite{Blokhuis:086101(2008)} a particular
model for the so-called \emph{surface density-density
correlation function} $S(\mathbf{q}):=(\triangle\rho)^{-2}\,\iint_{-\infty}^{+\infty}\, G(z,z',\mathbf{q})\mathrm{d}z\mathrm{d}z'$
(see Eq.\,(1) in Ref.\,\cite{Blokhuis:086101(2008)}
where $\triangle\rho:=\rho^{-}-\rho^{+}$ is the difference
of the coexisting bulk densities) is proposed in order
to explain the numerical data. $S_{\mathsf{BKV}}(\mathbf{q})$
in \cite{Blokhuis:086101(2008)} contains an interfacial
contribution $S_{hh}(\mathbf{q})$ and a bulk contribution
$S_{b}(\mathbf{q})$ (see Eq.\,(11) in \cite{Blokhuis:086101(2008)}).
Since $S_{b}(\mathbf{q})$ is modeled independently
from $S_{hh}(\mathbf{q})$ (the bulk density fluctuations
are considered to be \emph{uncorrelated} from the interface
fluctuations, see Eq.\,(10) in \cite{Blokhuis:086101(2008)})
and only $S_{hh}(\mathbf{q})$ is used to adjust the
data fit (Eqs.\,(14) and (15) in \cite{Blokhuis:086101(2008)}),
we focus here only on $S_{hh}(\mathbf{q})$. In order
to compare $S(\mathbf{q})$ resulting from Eq.\,\eqref{eq:G_planar_interface_Fourier_final}
to $S_{hh}(\mathbf{q})$, we adpot $v^{\mathsf{ext}}\equiv0$
and use also the Helfrich form $\eta(\mathbf{q})/q^{2}=\sigma+q^{2}\kappa$
(see Eqs.\,(8) and (9) in \cite{Blokhuis:086101(2008)}),
where $\sigma$ means the macroscopic surface tension
and $\kappa\geq0$ means the (phenomenological) bending
rigidity. From Eq.\,(4) in \cite{Blokhuis:086101(2008)}
we identify $\partial_{z}\rho_{\circ}(z)\hat{\lopds}_{\mathsf{BKV}}(z,\mathbf{q})=\frac{1}{2}\,\rho_{1}(z)\, q^{2}$,
where $\rho_{1}(z)$ is given by Eq.\,(26) in \cite{Blokhuis:086101(2008)}.
Interestingly, the kernels $\hat{\lopds}_{\mathsf{MD}}(z,\mathbf{q})$
and $\hat{\lopds}_{\mathsf{BVK}}(z,\mathbf{q})$ show
the same $q$-dependence, but the weight functions
$\rho_{H}(z)$ and $\rho_{1}(z)$, respectively, are
different. In particular, it is $\rho_{1}(z=0)\neq0$
so that the isodensity condition Eq.\,(\ref{eq:1})
is not fulfilled. By evaluating the integral $\frac{1}{\triangle\rho}\int\partial_{z}\rho_{\circ}(z)\,\big(1-\hat{\lopds}(z,\mathbf{q})\big)\,\mbox{d}z$,
we then obtain from Eq.\,\eqref{eq:G_planar_interface_Fourier_final}\begin{equation}
S(\mathbf{q})=\frac{(1+C\, q^{2})^{2}}{\beta\sigma q^{2}\,(1+\frac{\kappa}{\sigma}\, q^{2})}\label{eq:Surface_DD_corr_simple}\end{equation}
with $C=C_{H}\xi^{2}$ ($\mathsf{MD}$) or $C=0$ ($\mathsf{BKV}$),
respectively (see \cite{remark3} and Eq.\,(6) in
\cite{Blokhuis:086101(2008)}). The best fit to the
data in \cite{Blokhuis:086101(2008)} \emph{for in
the entire $q$-range} results from Eq.\,(15) in \cite{Blokhuis:086101(2008)}
which implies $S_{hh}(\mathbf{q})=(\beta\sigma q^{2})^{-1}\,(1+\frac{\kappa}{\sigma}\, q^{2})$
with $\kappa=-\kappa_{\mathsf{BKV}}>0$ because $\kappa_{\mathsf{BKV}}$
is found to be negative (Table\,I in \cite{Blokhuis:086101(2008)}).
This form for $S_{hh}(\mathbf{q})$ is obtained from
Eq.\,\eqref{eq:Surface_DD_corr_simple} for $\kappa=C\,\sigma$
if $C>0$ which is indeed the case for the $\hat{\lopds}_{\mathsf{MD}}(z,\mathbf{q})$
but not for $\hat{\lopds}_{\mathsf{BVK}}(z,\mathbf{q})$.
Since Eq.\,\eqref{eq:G_planar_interface_Fourier_final}
expresses the consistency between the height correlation
function and the density covariance function, this
result indicates that the kernel $\hat{\lopds}_{\mathsf{MD}}(z,\mathbf{q})$
seems to be even more consistent than $\hat{\lopds}_{\mathsf{BVK}}(z,\mathbf{q})$
with both, the assumed Helfrich form for $\eta(q)/q^{2}$
\emph{and} the numerical results for $S_{hh}(\mathbf{q})$
\emph{for the hole $q$-range} in \cite{Blokhuis:086101(2008)}.
In particular, the \emph{negative} bending rigitity
$\kappa_{\mathsf{BKV}}$ in \cite{Blokhuis:086101(2008)}
appears rather artificially from the truncated Taylor
expansion of $S_{hh}(\mathbf{q})$ in Eq.\,(14) in
\cite{Blokhuis:086101(2008)}, which is assumed to
be applicable for all $q$. Another main result of
\cite{Blokhuis:086101(2008)} is, that $\kappa_{\mathsf{BKV}}$
vanishes for $T\rightarrow T_{c}$. Using $\kappa=C_{H}\sigma\xi^{2}$
from the relation above, it then follows that $C_{H}$
vanishes with $T\rightarrow T_{c}$. This is in \emph{qualitative}
agreement with $C_{H}(T)\sim\xi^{-s}(T)$ for $s>0$
which has been concluded from general considerations
about the expected quantitative influence of $\hat{\lopds}_{\mathsf{MD}}(z,\mathbf{q})$
at higher temperatures (see comments on Eq.\,(55)
in \cite{Hiester:184701(2006)} and discussion on Eq.\,(4.72)
in \cite{mydiss}), but without determining the exponent
$s$ rigorously \cite{remark7}. Thus, although our
approach does not provide a particular theory for $\hat{\lopds}(z,\mathbf{q})$,
it might serve as a tool in order to verify the consistency
of various models with the underlying density covariance
function. 

X-ray experiments indicate $\hat{\lopds}(z;\mathbf{q})\sim q^{2}$
for isotropic interfaces \cite{Fradin:871(2000), Mora:216101(2003), Li:136102(2004), Luo:4527(2006), Li:235426(2006), Lin:106(2005)}.
But a separation of the height correlations from the
scattering data corresponds to the Eq.\,\eqref{eq:G_planar_interface_Fourier_final}.
As discussed above, any choice for $\hat{L}$ leads
to a height correlation function which is consistent
with $\hat{G}$. This makes probably clear uncertainty
of such procedures at larger $q$ values. Although
a generalization of Eqs.\,(\ref{eq:inv_height_corr})
to multi-component liquid mixtures as well as for spherical
interfaces can be performed, the more challenging task
is to identify the correlation ratios or the local
compressibility, respectively, hidden in $\lop$.

\begin{acknowledgments}
We have benefited from discussions with H. Leschke,
S. Dietrich, K. Mecke, M. Oettel, G. Schr\"oder-Turk,
and P. Tarazona. This work has been supported by the
Deutsche Forschungsgemeinschaft, Schwerpunkt Nanofluidik,
Grant ME1361/9-1.
\end{acknowledgments}


\begin{thebibliography}{34}
\bibitem{remark}
The related  mathematical problem can be outlined as follows.
For  a suitable subset \mbox{$M\subset\mathbb{R}^2$} we consider 
two random processes, \mbox{$\rd:\mathbb{R}^3\rightarrow \mathbb{R}$}
and \mbox{$Y: M\rightarrow \mathbb{R}^3$}. 
$\rd$ and $Y$ are related  by the (coordinate) projection 
\mbox{$P:\mathbb{R}^3\rightarrow M$ and the mean value  $\overline{X}$ of $\rd$}
via \mbox{$\rd(\mathbf{r}) = \overline{X}\big(\mathbf{r}+Y(P(\mathbf{r}))\big)$}.
If the covariance function of $\rd$ is given by
$G(\mathbf{r},\mathbf{r}')=\overline{\big(\rd(\mathbf{r})-\overline{X}(\mathbf{r})\big)
\big(\rd(\mathbf{r}')-\overline{X}(\mathbf{r}')\big)}$ the question is, 
what is the mean value and the covariance function of $Y$.
\bibitem{remark4}
An example of a function $f:\mathbb{R}\rightarrow\mathbb{R}$, which itself and its first derivative is  bounded but whose second derivative is unbounded, is $f(x)=(1-\mathrm{e}^{-x^4})\,\sin(\frac{1}{x^2})$. In particular, it is $f(0)=f'(0)=0$, $|f''(x\rightarrow 0)|\rightarrow\infty$ and the Fourier tranform of $f(x)$ exists since $|f(x)|$ is integrable.
\bibitem{remark5}
It is possible to allow mathematically for unbounded random densities $\rd(\mathbf{r})$. But from the physical point of view follows, that density configurations with $|\rd(\mathbf{r})|=\infty$ as well as the corresponding interface configurations occur with zero probability which results also in a limited $q$-range   of $\hat{\lopds}_{\mathsf{MD}}(z,\mathbf{q})$.
\bibitem{remark6}
If there is a number $c(z)\in\mathbb{R}^{+}$ so that  $|\lopds(z,\mathbf{R})|\leq c(z)$ for all $\mathbf{R}\in\mathbb{R}^2$ and $|z|<\infty$ then 
$\Vert\lop(z)u\Vert_1\leq c(z)\,\Vert u\Vert_1<\infty$.
\end{thebibliography}
\end{document}